\documentclass[aps,prl,reprint,superscriptaddress,longbibliography,floatfix]{revtex4-2}

\usepackage{amsmath,amssymb,bm,mathtools}
\usepackage{graphicx}
\usepackage{microtype}
\usepackage{xcolor}
\usepackage{hyperref}
\usepackage{booktabs}
\usepackage{multirow}
\usepackage{longtable}
\hypersetup{colorlinks=true,citecolor=blue,linkcolor=blue,urlcolor=blue}
\graphicspath{{figures/}}
\newcommand{\Prob}{\mathbb P}
\newcommand{\ind}{\mathbf 1}
\newcommand{\safefigure}[3]{%
\IfFileExists{#2}{\includegraphics[#1]{#2}}{%
\fbox{\parbox[c][#3][c]{0.94\linewidth}{\centering
\textbf{Production figure to be supplied}\\[2pt]
\texttt{\detokenize{#2}}}}}}

\makeatletter
\let\SI@affiliation\affiliation
\let\SI@email\email
\let\SI@thanks\thanks
\makeatother

\begin{document}

\title{Emergent Equilibrium Structure Along a Critical Cluster Recursion}

\author{Shuo Wei}\email{shuowei@mail.ustc.edu.cn}
\affiliation{
Department of Modern Physics, University of Science and Technology of China, Hefei 230026, China}
\author{Abbas Ali Saberi}
\email{asaberi@constructor.university}
\affiliation{School of Science, Constructor University, Campus Ring 1, 28759 Bremen, Germany}
\author{Youjin Deng}
\email{yjdeng@ustc.edu.cn}
\affiliation{
Department of Modern Physics, University of Science and Technology of China, Hefei 230026, China}
\affiliation{Hefei National Laboratory, University of Science and Technology of China, Hefei 230088, China}
\affiliation{Hefei National Research Center for Physical Sciences at the Microscale and School of Physical Sciences, University of Science and Technology of China, Hefei 230026, China}
\affiliation{College of Physics, Guizhou University, Guiyang 550025, China}
\date{\today}

\begin{abstract}
Critical universality does not determine the microscopic conditional structure of a probability measure. We study a bicolored cluster recursion constrained to remain critical at every generation, with no equilibrium spin measure or fixed coupling imposed. In both two and three dimensions, the resulting history-dependent sequence develops a common Ising/Fortuin--Kasteleyn (FK) compatibility structure: the second-shell dependence of a one-site conditional law is strongly suppressed, nearest-neighbor effective couplings move progressively toward one another near the critical Ising value, and cluster and interface observables organize around the corresponding FK geometry. An exact cluster-coloring factorization singles out $q=2$ as the point where the residual connectivity weight disappears from the two-color spin marginal. Thus equilibrium-compatible conditional structure can emerge along a trajectory that remains critical throughout.
\end{abstract}

\maketitle

Universality is powerful precisely because critical behavior forgets microscopic detail.  Distinct equilibrium models flow to the same long-distance fixed point, and even nonequilibrium dynamics that violate detailed balance can display equilibrium critical scaling or asymptotic thermalization \cite{Kadanoff1966,WilsonKogut1974,Fisher1998, Cardy1996, Grinstein1985,Risler2005,Sieberer2013,Sieberer2025}.  Yet an equilibrium probability measure contains information that universality does not fix.  A Gibbs state contains more information than its scaling dimensions and correlation functions: its local conditional probabilities are compatible with a local interaction. In particular, once the configuration outside a finite region is given, that interaction specifies the probability law inside the region \cite{Dobrushin1968,LanfordRuelle1969}. This local structure is not automatically preserved under a transformation. Indeed, even coarse-graining maps intended to retain long-wavelength behavior can map Gibbs measures to non-Gibbsian ones, whose local conditional probabilities no longer admit such a well-behaved local description \cite{Schonmann1989,vanEnterFernandezSokal1993}. This raises the converse question addressed here: \emph{can a transformation constrained only to remain critical progressively organize its microscopic conditional probabilities toward a local equilibrium structure?}

\begin{figure*}[t]
\centering
\safefigure{width=0.98\textwidth}{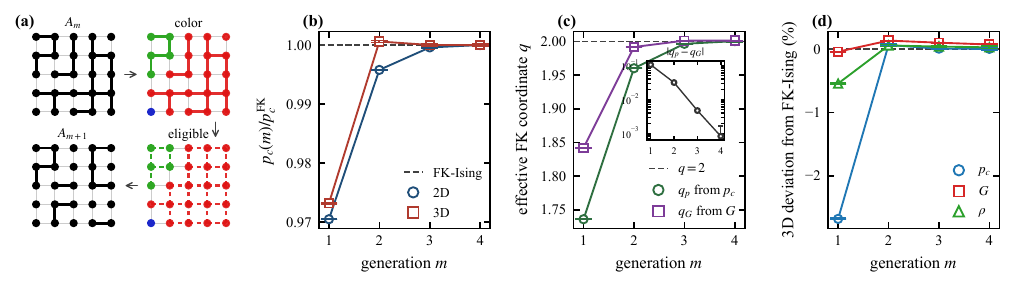}{5.8cm}
\caption{\textbf{Critical trajectory and local FK closure.}
(a) Recursive construction: clusters of the critical configuration $A_m$ are independently colored by $\sigma=\pm1$, and equal-spin nearest neighbors are rebonded with the generation-dependent critical probability $p_c(m+1)$.
(b) Critical bond-insertion thresholds normalized by the corresponding FK--Ising values, $p_c(m)/p_c^{\rm FK}$, for $m=1,\ldots,4$ in two and three dimensions. The horizontal line at unity marks the respective FK--Ising threshold.
(c) Independent 2D projections $q_p$ and $q_G$ obtained from $p_c$ and $G$, respectively, with $q=2$ as reference; inset: $|q_p-q_G|$ versus generation on a logarithmic scale.
(d) Relative deviations of the 3D local observables $p_c$, $G$, and $\rho$ from their FK--Ising reference values.}
\label{fig:trajectory}
\end{figure*}
\paragraph{Critical-to-critical recursion.}

Percolation provides a paradigmatic setting in which criticality is encoded directly in random connectivity and fractal geometry \cite{Saberi2015}. Cluster representations provide an unusually direct setting in which to examine the relation between connectivity and local probability structure. The Fortuin--Kasteleyn (FK) random-cluster representation and the Edwards--Sokal coupling establish an exact correspondence between $q$-state Potts Gibbs measures, including the Ising model at $q=2$, and random-cluster configurations \cite{FortuinKasteleyn1972,EdwardsSokal1988}. By contrast, the map obtained by coloring connected components and then hiding the cluster variables is nonlocal: information can be transmitted through clusters on arbitrarily large scales. The resulting divide-and-color or fuzzy-Potts measures can be Gibbsian or non-Gibbsian depending on the underlying ensemble and parameters \cite{Haggstrom2001,Haggstrom2003,Balint2010}.  At the same time, recursive and iterative percolation reveal a different phenomenon: critical cluster configurations can be repeatedly reorganized while scale invariance survives, producing generation-dependent critical geometries rather than returning to a single fixed ensemble \cite{LiuDengJacobsen2015,LiDeng2024,WeiLiuSunDengLi2026}.  This motivates the question of how the local conditional structure evolves along this sequence of critically tuned ensembles.

We address this question with a bicolored recursion that maps one critical bond ensemble to another. At each generation, inherited clusters are independently assigned spins $\sigma=\pm1$, and equal-spin nearest neighbors are rebonded with a single probability tuned only to restore criticality. For these binary color variables, the nearest-neighbor Ising model provides the natural equilibrium reference. Unlike the Swendsen--Wang Markov chain, whose stationary distribution is the Gibbs measure at a fixed chosen coupling \cite{SwendsenWang1987,EdwardsSokal1988}, or invaded-cluster algorithms, which use a prescribed equilibrium cluster representation to locate its critical point without prior knowledge of the critical coupling \cite{Machta1995,Moriarty1999}, the recursion has no imposed Hamiltonian, fixed equilibrium coupling, or stationary target distribution. It instead defines a history-dependent sequence of critically tuned ensembles. Any approach of the recursively generated measures toward an Ising-compatible conditional structure is therefore an outcome of the critically retuned trajectory rather than a prescribed stationary property.

Our numerical results show that conditional statistics and critical geometry move toward the same FK--Ising reference structure. An exact cluster-coloring factorization singles out $q=2$: for equal-probability two-coloring of a finite random-cluster configuration, it is the unique value at which the residual cluster factor disappears from the two-color spin marginal, leaving a local Ising factor. The recursively generated ensembles are not assumed to remain within that family. Nevertheless, three distinct diagnostics move toward this reference point: the second-shell dependence is strongly suppressed, effective couplings inferred separately from different nearest-neighbor fields move progressively closer to one another near the critical Ising value, and cluster and interface observables move towards their FK--Ising values, while the red-bond
 and backbone sectors retain larger finite-generation deviations. Thus evolving criticality extends beyond scaling geometry: along the same critically tuned sequence, local conditional structure and global connectivity acquire multiple signatures of the same FK--Ising reference structure.

Let $A_m$ denote the occupied nearest-neighbor bonds at generation $m$ on either a periodic square or simple-cubic lattice, with $A_0$ ordinary critical bond percolation. Each connected component of $A_m$ is independently assigned a sign $\sigma=\pm1$, and equal-spin nearest-neighbor pairs are rebonded according to
\begin{equation}
\Prob(\langle ij\rangle\in A_{m+1}\mid \sigma)
=p_{m+1}\ind_{{\sigma_i=\sigma_j}} .
\label{eq:update}
\end{equation}
The probabilities defining all preceding generations are held at their previously determined critical values, while $p_{m+1}$ alone is varied to locate the next transition. Finite-size wrapping analysis yields a nontrivial critical point at every investigated generation in both dimensions; the corresponding thresholds are listed in Table~\ref{tab:summary}. The thresholds change strongly at first and then approach the FK--Ising values: in 2D, for example, $p_c(m)$ moves from $0.56850(2)$ at $m=1$ to $0.58581(5)$ at $m=4$, compared with $2-\sqrt2=0.585786\ldots$. The recursion therefore traces a sequence of distinct critically tuned ensembles rather than repeatedly sampling one fixed critical ensemble. Details of the simulation protocol, finite-size analyses, extended data, and conditional-locality measurements are provided in the Supplemental Material \cite{SupplementalMaterial}.

The construction yields an exact bond-level relation. If $G_m$ is the probability that two nearest neighbors are connected in $A_m$, their colors agree with probability $(1+G_m)/2$: connected sites agree identically, while disconnected clusters receive the same color with probability $1/2$. Hence
\begin{equation}
\rho_{m+1}=p_c(m+1)\frac{1+G_m}{2},
\label{eq:rho}
\end{equation}
where $\rho_{m+1}$ is the occupied-bond density. In 2D, the measured values of $G_m$ and $\rho_m$ move toward $1/\sqrt2$ and $1/2$, respectively; together with Eq.~\eqref{eq:rho}, this trend is consistent with $p_*=2-\sqrt2$. On the square lattice, exact FK critical relations allow the critical threshold and nearest-neighbor connectivity to be converted independently into effective FK coordinates \cite{Baxter1973,BaxterKellandWu1976}:
$q_p(m)=[p_c(m)/(1-p_c(m))]^2$ and
$q_G(m)=[2(1-G_m)/(2G_m-1)]^2$.
They evolve from $(q_p,q_G)=(1.7358(3),1.8418(1))$ at $m=1$ to $(2.0004(8),2.0012(1))$ at $m=4$, while $|q_p-q_G|$ falls from $0.1060(3)$ to $0.0008(8)$. Thus, when interpreted through the exact FK critical relations, the two observables yield separately inferred effective FK coordinates that move toward the same reference value, $q=2$. This comparison is a projection onto the exactly known FK critical family; it does not assume that the finite-generation recursive measures themselves belong to that family. In 3D, where no analogous exact one-parameter critical line is available, $p_c$, $G$, and $\rho$ show a corresponding joint movement toward their independently known FK--Ising references.

\begin{figure*}[t]
\centering
\safefigure{width=0.98\textwidth}{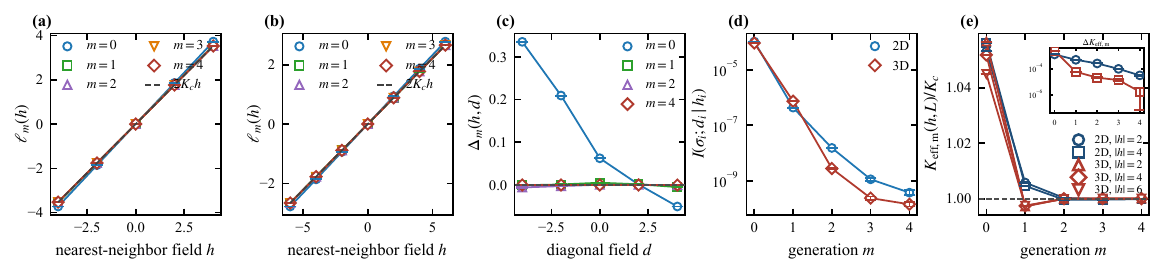}{6.7cm}
\caption{\textbf{Conditional approach to the Ising local rule.}
(a) $\ell_m(h)$ in 2D at $L=1024$ for $m=0,\ldots,4$; the reference line is $2K_c h$.
(b) Corresponding 3D results at $L=128$.
(c) Second-shell dependence $\Delta_m(h,d)=\ell_m(h,d)-\ell_m(h)$ in 2D at $L=1024$, for $h=2$ and generations $m=0,1,2,4$, shown on a common vertical scale.
(d) Conditional mutual information $I_m(\sigma_i;d_i\mid h_i)$ versus generation at $L=1024$ in 2D and $L=128$ in 3D; the late-generation values approach the numerical resolution floor of the aggregate-count estimator.
(e) Effective couplings $K_{\mathrm{eff},m}(h,L)/K_c$ at $L=1024$ in 2D and $L=128$ in 3D. Inset: $\Delta K_{\mathrm{eff},m}(L)=\max_{|h|}K_{\mathrm{eff},m}(|h|,L)-\min_{|h|}K_{\mathrm{eff},m}(|h|,L)$ versus generation on a logarithmic scale.}
\label{fig:locality}
\end{figure*}

\paragraph{Why Ising is the benchmark.}
The binary cluster colors make the nearest-neighbor Ising model the natural equilibrium reference. On a finite graph $G$, write
\begin{equation}
\mathcal Z_{\rm RC}(G;p,q)=
\sum_{B\subseteq E(G)}p^{|B|}(1-p)^{|E(G)|-|B|}q^{C_G(B)},
\label{eq:ZRC}
\end{equation}
where $C_G(B)$ is the number of connected components of the spanning subgraph $(V(G),B)$. If the input is a random-cluster configuration with bond probability $p$ and cluster weight $q$, and each bond cluster is independently assigned $\sigma=\pm1$, its spin marginal is
\begin{equation}
\nu_{p,q}(\sigma)
\propto
\exp\!\left[
K(p)\sum_{\langle ij\rangle}\sigma_i\sigma_j
\right]
\mathcal Z_{\rm RC}
\!\left(
G_\sigma;p,\frac{q}{2}
\right),
\label{eq:factorization}
\end{equation}
where $K(p)=-\frac{1}{2}\ln(1-p)$ and $G_\sigma$ is the spanning subgraph of equal-spin nearest-neighbor edges. The second factor retains the cluster dependence. At $q=2$, however, $\mathcal Z_{\rm RC}(G_\sigma;p,1)=1$, and the colored spins are exactly the nearest-neighbor Ising model at coupling $K(p)$: the Edwards--Sokal correspondence \cite{EdwardsSokal1988}.

This is a benchmark, not an assumption about the recursive ensembles. To state the distinction exactly, let $\phi_{p,2}$ be the $q=2$ FK measure, and let $f=d\mu/d\phi_{p,2}$ be the density of an arbitrary bond law $\mu$ on the same finite graph. Its two-color spin marginal obeys
\begin{equation}
\nu_{\mu}(\sigma)
=
\nu^{\rm Ising}_{K(p)}(\sigma)\,
\bigl\langle f(A)\mid\sigma\bigr\rangle_{p,2}.
\label{eq:general_factorization}
\end{equation}
Here $\nu^{\rm Ising}_{K(p)}(\sigma)$ is the normalized nearest-neighbor Ising probability at coupling $K(p)$, and $\langle f(A)\mid\sigma\rangle_{p,2}$ is the Edwards--Sokal conditional average of the relative bond weight $f(A)$ over configurations compatible with $\sigma$. Consequently, matching selected FK reference quantities alone does not establish equality with the Ising law. We instead test its local consequences: a nearest-neighbor Ising conditional law has neither a next-nearest-neighbor dependence nor an effective coupling that varies with the nearest-neighbor field.

\begin{table*}[t]
\caption{\textbf{Selected critical and geometric diagnostics.} Values are shown for all simulated generations together with the corresponding Ising/FK--Ising references. Parentheses denote statistical fit uncertainties; propagated uncertainty from earlier-generation critical thresholds is not included.}
\label{tab:summary}
\centering
\begingroup
\setlength{\tabcolsep}{4.2pt}
\renewcommand{\arraystretch}{1.05}
\begin{tabular*}{\textwidth}{@{\extracolsep{\fill}}lllllll@{}}
\toprule
Dimension & Observable & $m=1$ & $m=2$ & $m=3$ & $m=4$ & Ising/FK--Ising \\
\midrule
\multirow{6}{*}{2D}
& $p_{\mathrm c}$ & 0.56850(2) & 0.58334(3) & 0.58557(3) & 0.58581(5) & $2-\sqrt2$ \\
& $G$ & 0.712124(4) & 0.707360(4) & 0.707072(4) & 0.707070(4) & $1/\sqrt2$ \\
& $d_{\mathrm{FK}}$ & 1.8788(3) & 1.8755(4) & 1.8750(6) & 1.875(1) & $15/8$ \\
& $d_{\mathrm{spin}}$ & 1.9494(4) & 1.9484(4) & 1.9480(3) & 1.9480(5) & $187/96$ \\
& $d_{\mathrm{red}}$ & 0.605(2) & 0.567(2) & 0.556(6) & 0.554(2) & $13/24$ \\
& $d_{\mathrm B}$ & 1.709(1) & 1.724(2) & 1.728(2) & 1.729(2) & $1.7321\ldots$ \\
\addlinespace[2pt]
\multirow{5}{*}{3D}
& $p_{\mathrm c}$ & 0.34849(3) & 0.35832(6) & 0.35811(5) & 0.35810(7) & 0.358091335(6) \\
& $G$ & 0.330066(4) & 0.330644(4) & 0.330534(4) & 0.33044(2) & 0.330204(2) \\
& $d_{\mathrm{FK}}$ & 2.482(1) & 2.484(3) & 2.479(3) & 2.477(3) & 2.4818511(10) \\
& $d_{\mathrm{red}}$ & 0.882(5) & 0.836(3) & 0.835(4) & 0.826(5) & 0.757(2) \\
& $d_{\mathrm B}$ & 2.077(3) & 2.107(3) & 2.110(6) & 2.107(3) & 2.1673(15) \\
\bottomrule
\end{tabular*}
\endgroup
\end{table*}
\paragraph{Approach to the local Ising rule.}

We test this directly in the colored configurations before rebonding. For each site, let $h_i=\sum_{j\in {\rm NN}(i)}\sigma_j$ be the nearest-neighbor spin sum and let $d_i=\sum_{j\in {\rm NNN}(i)}\sigma_j$ denote a second-shell field, formed from the four diagonal neighbors in 2D and the twelve face-diagonal neighbors in 3D. We measure
\begin{equation}
\ell_m(h,d)=\log\frac{P_m(\sigma_i=+1\mid h_i=h,d_i=d)}
{P_m(\sigma_i=-1\mid h_i=h,d_i=d)} .
\label{eq:logodds}
\end{equation}
A nearest-neighbor Ising specification gives $\ell(h,d)=2Kh$, independent of $d$. We also define the conditional log-probability ratio given only the nearest-neighbor field,
\begin{equation}
\ell_m(h)=\log\frac{P_m(\sigma_i=+1\mid h_i=h)}
{P_m(\sigma_i=-1\mid h_i=h)}.
\label{eq:logodds_h}
\end{equation} The measurement therefore separates dependence on the measured second-shell field,
$\Delta_m(h,d)=\ell_m(h,d)-\ell_m(h)$,
from deviations from a single nearest-neighbor coupling. For each nonzero value of the nearest-neighbor field $h$, we define
\begin{equation}
K_{\mathrm{eff},m}(h,L)=\frac{\ell_m(h,L)}{2h}.
\label{eq:keff}
\end{equation}
We summarize the residual field dependence by
\[
\Delta K_{\mathrm{eff},m}(L)
=
\max_{|h|}K_{\mathrm{eff},m}(|h|,L)
-
\min_{|h|}K_{\mathrm{eff},m}(|h|,L).
\]
We also measure the conditional mutual information
\begin{equation}
I_m(\sigma_i;d_i\mid h_i)
=
\sum_{\sigma,d,h}P_m(\sigma,d,h)
\log\frac{P_m(\sigma,d\mid h)}
{P_m(\sigma\mid h)P_m(d\mid h)},
\label{eq:cmi}
\end{equation}
which vanishes when $\sigma_i$ and $d_i$ are conditionally independent given $h_i$. Here $L$ is the linear system size. An exact nearest-neighbor Ising conditional law requires both $\Delta_m(h,d)=0$ and an $h$-independent $K_{\mathrm{eff},m}(h,L)$.

Figure~\ref{fig:locality} summarizes these local diagnostics across the investigated generations. At $m=0$, $\Delta_m(h,d)$ has a clear systematic dependence on the second shell, and the values of $K_{\mathrm{eff},m}$ obtained at different $h$ are visibly separated. Most of the resolved shell dependence is strongly suppressed after the first iteration; the conditional mutual information $I_m(\sigma_i;d_i\mid h_i)$ shows the same suppression. Increasing the recursion generation also strongly suppresses the spread among the effective couplings extracted at different $h$. At $L=1024$ in 2D, the $|h|=2,4$ estimates evolve from $0.464083(4)$ and $0.465448(2)$ at $m=0$ to $0.440673(3)$ and $0.440640(2)$ at $m=4$, compared with $K_c=0.44068679\ldots$. At $L=128$ in 3D, the $|h|=2,4,6$ estimates evolve from $0.234195(1)$, $0.233130(1)$, and $0.231614(1)$ to $0.221674(2)$, $0.221672(1)$, and $0.221672(1)$, compared with $K_c=0.221654626$ \cite{Ferrenberg2018}. Over the accessible size range, increasing $L$ at fixed generation does not make the effective couplings extracted at different $h$ coincide, whereas increasing the recursion generation strongly reduces their separation. Within this finite-shell diagnostic, the measured dependence on $d_i$ is suppressed and the effective couplings approach a single-coupling Ising form.

\begin{figure*}[t]
\centering
\safefigure{width=0.98\textwidth}{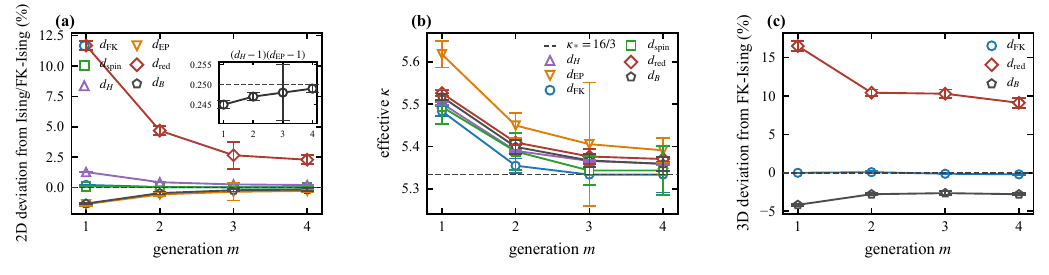}{6.5cm}
\caption{\textbf{Common FK--Ising structure across geometric scales.}
(a) Relative deviations of the 2D FK-cluster mass, geometric spin-cluster mass, hull, external perimeter, red-bond, and backbone dimensions from their Ising/FK--Ising values for $m=1,\ldots,4$, with the reported fit uncertainties. Inset: $(d_H-1)(d_{\rm EP}-1)$ with reference value $1/4$.
(b) A posteriori effective FK/Coulomb-gas coordinates $\kappa$ obtained by inserting each measured dimension separately into the corresponding reference relation. No Coulomb-gas description of the recursive measures is assumed. The hull, external-perimeter, FK-cluster, and red-bond inversions use the exact critical FK/Potts relations; the spin and backbone inversions use the equal-color one-arm and backbone relations, respectively \cite{LiuSunYuZhuang2024,NolinQianSunZhuang2024}. The horizontal line denotes $\kappa_*=16/3$. The enhanced red-bond sensitivity at Ising, $\partial_\kappa d_{\rm red}/\partial_\kappa d_{\rm FK}=43/3$, is consistent with its larger visible displacement.
(c) Relative deviations of the 3D FK-cluster, red-bond, and backbone dimensions from the numerical FK--Ising references. The panel displays the hierarchy of finite-generation corrections; no common one-parameter geometric description is assumed in 3D.}
\label{fig:geometry}
\end{figure*}

\paragraph{
Common FK--Ising geometric organization across scales.}
Geometric observables provide an independent test of the apparent FK--Ising organization. In 2D, the FK-cluster mass, geometric spin-cluster mass, hull, external-perimeter, red-bond, and backbone dimensions all move toward their corresponding Ising/FK--Ising values \cite{SaleurDuplantier1987,AharonyAsikainen2003,DengBloteNienhuis2004,DengBloteNienhuisBackbone2004}. The hull and external perimeter furnish an especially simple check of the Duplantier hull--external-perimeter duality \cite{Duplantier2004}: their measured combination $(d_H-1)(d_{\rm EP}-1)$ evolves from $0.245(1)$ to $0.249(1)$, approaching the FK--Ising value $1/4$. Selected numerical values across the trajectory are collected in Table~\ref{tab:summary}.

The unequal rates of change can be compared, a posteriori, with the exact two-dimensional reference relations. We invert the appropriate relation for each measured dimension separately to define an effective $\kappa$; this comparison does not assume that the recursive ensembles themselves obey a Coulomb-gas description. At later generations, the separately inferred values lie closer to the common FK--Ising reference $\kappa_*=16/3$ [Fig.~\ref{fig:geometry}(b)] \cite{SaleurDuplantier1987,AharonyAsikainen2003,DengBloteNienhuis2004,LiuSunYuZhuang2024,NolinQianSunZhuang2024}. The red-bond sector is particularly sensitive in this projection: at $\kappa_*=16/3$,
\[
\frac{\partial_\kappa d_{\rm red}}{\partial_\kappa d_{\rm FK}}=\frac{43}{3}.
\]
Thus, a small effective-$\kappa$ displacement produces a much larger visible deviation in the red-bond dimension than in the leading cluster mass.

Three dimensions provide a complementary comparison because no exact one-parameter Coulomb-gas description is available. The data instead show a hierarchy of finite-generation deviations. The critical threshold, nearest-neighbor connectivity, bond density, and leading FK-cluster mass lie comparatively close to their FK--Ising reference values, whereas the red-bond and backbone observables retain larger deviations \cite{Hou2019,DengBlote2004Red3D}. At $m=4$, for example, $d_{\rm red}=0.826(5)$ compared with $0.757(2)$, and $d_B=2.107(3)$ compared with $2.1673(15)$, whereas $d_{\rm FK}=2.477(3)$ compared with $2.4818511(10)$. In both dimensions, the results suggest a common ordering: local conditional diagnostics and leading geometric quantities approach their reference values over fewer recursion generations than the red-bond and backbone sectors.

\textit{Conclusions.---} The recursion consequently separates three structures that coincide in an equilibrium critical model: infrared criticality, local equilibrium-compatible conditional structure, and the associated connectivity geometry. Criticality is imposed at every generation, whereas the local conditional structure and connectivity geometry are not and evolve with generations. The exact $q=2$ cancellation in the random-cluster factorization identifies why FK--Ising is the natural microscopic reference: for equal-probability two-coloring of a $q$-random-cluster configuration, only $q=2$ makes the residual cluster factor identically unity in the spin marginal. The conditional and geometric measurements show that the recursively generated critical sequence progressively develops signatures of this same compatibility structure through distinct local, bulk, interface, and connectivity diagnostics. 

These results suggest a broader dynamical view of critically tuned ensembles. At finite generation, the recursion traces a trajectory in the measured local and geometric observables, while its behavior in the joint large-generation and thermodynamic limit remains open. An important next step is to determine whether these limits commute and whether the generation map can be linearized near the observed FK--Ising-compatible regime. Varying the number of colors, spatial dimension, and underlying graph may reveal when repeated critical transformations suppress inherited connectivity sufficiently to develop an equilibrium-compatible local structure. More generally, our results suggest that critical transformations may be characterized not only by the infrared fixed points they approach or preserve, but also by the microscopic information they erase or retain while remaining critical.

\textit{Acknowledgments.---} S.W. and Y.D. acknowledge the support by the National Natural Science Foundation of China (NSFC) under Grant No. 12275263, as well as Quantum Science
and Technology-National Science and Technology Major Project (under Grant No. 2021ZD0301900). A.A.S. was funded by the Deutsche Forschungsgemeinschaft (DFG, German Research Foundation) under Project No.~557852701.

\textit{Data availability.---} The data and simulation and analysis codes that support the findings of this article are not publicly available upon publication because it is not technically feasible and/or the cost of preparing, depositing, and hosting them would be prohibitive within the terms of this research project. They are available from the authors upon reasonable request.

\makeatletter
\let\auto@bib@innerbib\@empty
\makeatother

\clearpage
\setcounter{equation}{0}
\setcounter{figure}{0}
\setcounter{table}{0}
\setcounter{section}{0}
\setcounter{page}{1}

\newcommand{\pc}{p_{\mathrm c}}
\newcommand{\placeholder}[1]{\textcolor{red}{\textbf{[To be completed: #1]}}}
\newcommand{\revision}[1]{#1}

\makeatletter
\frontmatter@init
\let\title\frontmatter@title
\let\author\frontmatter@author
\let\date\frontmatter@date
\let\maketitle\frontmatter@maketitle
\let\affiliation\SI@affiliation
\let\email\SI@email
\let\thanks\SI@thanks
\let\SI@origlabel\label
\renewcommand{\label}[1]{%
  \edef\SI@tempa{#1}%
  \def\SI@tempb{FirstPage}%
  \def\SI@tempc{LastBibItem}%
  \ifx\SI@tempa\SI@tempb
    \SI@origlabel{SuppFirstPage}%
  \else
    \ifx\SI@tempa\SI@tempc
      \SI@origlabel{SuppLastBibItem}%
    \else
      \SI@origlabel{#1}%
    \fi
  \fi}
\makeatother

\let\SIorigcite\cite
\let\SIorigbibitem\bibitem
\renewcommand{\cite}[1]{\SIorigcite{supp:#1}}
\renewcommand{\bibitem}[1]{\SIorigbibitem{supp:#1}}

\title{Supplemental Material for ``Emergent Equilibrium Structure Along a Critical Cluster Recursion''}

\author{Shuo Wei}
\affiliation{Department of Modern Physics, University of Science and Technology of China, Hefei 230026, China}
\author{Abbas Ali Saberi}
\affiliation{School of Science, Constructor University, Campus Ring 1, 28759 Bremen, Germany}
\author{Youjin Deng}
\affiliation{Department of Modern Physics, University of Science and Technology of China, Hefei 230026, China}
\affiliation{Hefei National Laboratory, University of Science and Technology of China, Hefei 230088, China}
\affiliation{Hefei National Research Center for Physical Sciences at the Microscale and School of Physical Sciences, University of Science and Technology of China, Hefei 230026, China}
\affiliation{College of Physics, Guizhou University, Guiyang 550025, China}
\date{\today}
\maketitle

\section{Simulation protocol}
\label{sec:simulation_protocol}

We simulate the following bicolored recursive procedure on periodic square and simple-cubic lattices.  The bond configuration at generation $m$ is denoted by $A_m$.  Generation zero is ordinary critical bond percolation.  To construct $A_{m+1}$, every bond cluster of $A_m$ is assigned an independent auxiliary spin $\sigma=\pm1$ with equal probabilities.  Each nearest-neighbor lattice edge $\langle ij\rangle$ is then independently included in the new configuration $A_{m+1}$ with probability
\begin{equation}
\Prob\bigl(\langle ij\rangle\in A_{m+1}\mid\sigma\bigr)
 =p_{m+1}\ind_{\{\sigma_i=\sigma_j\}}.
 \label{eq:supp_update}
\end{equation}
The auxiliary spins are discarded after the bond-insertion step.  The connected components of the resulting configuration define the bond clusters to be colored at the following generation.

All systems use periodic boundary conditions.  In two dimensions, the simulated linear sizes are
\begin{equation}
 L=16,32,\ldots,2048,
 \label{eq:supp_sizes_2d}
\end{equation}
and in three dimensions they are
\begin{equation}
 L=16,24,32,48,64,96,128,192.
 \label{eq:supp_sizes_3d}
\end{equation}
For each point $(p,L)$ in a wrapping-probability scan, we generate $10^6$ independent realizations.  A realization includes the entire sequence of preceding generations; it is not obtained by continuing a previously sampled configuration.  Thus the quoted statistical uncertainties of the wrapping probabilities are based on independent recursive histories.

The critical sequence is generated sequentially.  When locating the transition at generation $m$, all earlier probabilities are held at their adopted critical values,
\begin{equation}
 A_0 \xrightarrow{\pc(1)} A_1 \xrightarrow{\pc(2)} \cdots
 \xrightarrow{\pc(m-1)} A_{m-1} \xrightarrow{p} A_m,
 \label{eq:supp_sequential}
\end{equation}
and only the final probability $p$ is varied.  The value at which $A_m$ is critical is denoted by $\pc(m)$.  This procedure is repeated separately for every generation studied.  In particular, the simulations do not iterate a fixed bond probability toward a stationary ensemble.

Bond clusters are identified after every bond-insertion step, and each cluster is assigned a single new auxiliary spin before the next step.

For a sequentially generated observable, the statistical error from the final sampling stage is distinct from possible systematic uncertainty arising from the adopted values of $\pc(j<m)$.  All uncertainties reported here and in the Letter are statistical fit uncertainties.  \revision{We do not presently have an efficient procedure for propagating the uncertainty in the earlier-generation thresholds through the recursive construction.}  The quoted error bars should therefore be regarded as lower bounds on the total uncertainty; the unquantified propagated contribution can make the actual uncertainty larger.

\section{Location of the critical points from wrapping probabilities}
\label{sec:wrapping_analysis}

For two-dimensional systems, we use $R_2(p,L)$, the probability that at least one bond cluster wraps around the torus in two independent directions.  For three-dimensional systems, we use the one-direction wrapping probability averaged over the three Cartesian directions,
\begin{equation}
 R_x(p,L)=\frac{R^{(x)}(p,L)+R^{(y)}(p,L)+R^{(z)}(p,L)}{3},
 \label{eq:supp_rx}
\end{equation}
where $R^{(\alpha)}$ is the probability that a bond cluster wraps in direction $\alpha$.  We write either observable as $R(p,L)$ when the same finite-size form is being used in both dimensions.

For each generation, curves for different $L$ cross in a narrow interval of $p$.  We fit the wrapping data in the neighborhood of the crossing to
\begin{align}
 R(p,L)={}&R_c+\sum_{k=1}^{2}a_k\bigl[(p-\pc)L^{y_p}\bigr]^k
 +b_1L^{-y_1}
 \nonumber\\
 &+c_1(p-\pc)L^{y_p-y_1}.
 \label{eq:supp_wrapping_fit}
\end{align}
Here $R_c$ is the critical wrapping probability for the chosen wrapping observable, $y_p$ controls the response to the final bond-insertion probability, and $y_1>0$ parameterizes the leading correction to scaling.  The fit is performed jointly over the selected values of $p$ and $L$.  \revision{The coefficient $c_1$ accommodates the leading correction to the slope.}

\revision{The exponent $y_p$ is associated with pivotal bonds (red bonds) for the final insertion step.}  A pivotal edge is one whose state changes the wrapping event.  An occupied pivotal edge whose removal disconnects the wrapping structure is a red bond.

To determine the final thresholds, we progressively increase the minimum size $L_{\min}$ retained in the fit and identify a range in which the estimate of $\pc$ is stable within its statistical uncertainty.  Tables~\ref{tab:wrap_2d} and \ref{tab:wrap_3d} show representative fits using the full form of Eq.~\eqref{eq:supp_wrapping_fit}; the threshold values adopted in the Letter are chosen from the corresponding stable ranges.  The tables are intended as a stability record rather than as an independent estimate of a systematic error.

\begin{table*}[t]
\caption{Representative two-dimensional wrapping-probability fits based on $R_2$ and the full form of Eq.~\eqref{eq:supp_wrapping_fit}.  Increasing $L_{\min}$ tests the stability of the threshold estimate.  The values adopted in the Letter are $\pc=0.56850(2)$, $0.58334(3)$, $0.58557(3)$, and $0.58581(5)$ for $m=1,2,3,$ and $4$, respectively.}
\label{tab:wrap_2d}
\begin{ruledtabular}
\scriptsize
\setlength{\tabcolsep}{2pt}
\renewcommand{\arraystretch}{0.92}
\begin{tabular}{lllllllllll}
 $m$ & $L_{\min}$ & $R_c$ & $a_1$ & $\pc$ & $y_p$ & $a_2$ & $b_1$ & $y_1$ & $c_1$ & $\chi^2/\mathrm{DOF}$ \\
 \colrule
 1 & 16 & 0.3965(6) & 1.3(1) & 0.56850(1) & 0.589(15) & -1.2(9) & 0.7(2) & 1.34(9) & 0.4(6.2) & 0.72 \\
   & 32 & 0.396(1)  & 1.3(2) & 0.56849(2) & 0.60(2)   & -0.9(9) & 0.4(3) & 1.1(3)  & 3.9(6.5) & 0.83 \\
   & 64 & 0.397(1)  & 1.3(2) & 0.56850(2) & 0.59(2)   & -1.2(1.0) & 6(32) & 1.9(1.4) & -30(315) & 0.45 \\
 2 & 16 & 0.416(2)  & 1.5(3) & 0.58333(1) & 0.55(2)   & -0.6(7) & 0.029(4) & 0.4(1) & -1.0(9) & 0.82 \\
   & 32 & 0.416(3)  & 1.4(3) & 0.58333(2) & 0.56(3)   & -0.5(6) & 0.03(1)  & 0.4(2) & -1.0(1.2) & 0.81 \\
   & 64 & 0.418(2)  & 1.3(2) & 0.58334(2) & 0.57(2)   & -0.5(6) & 0.06(9)  & 0.7(5) & -1.5(3.3) & 0.78 \\
   & 128& 0.418(3)  & 1.5(6) & 0.58334(2) & 0.55(5)   & -0.6(8) & 0.1(7)   & 0.9(1.3) & -12(56) & 0.88 \\
 3 & 32 & 0.423(2)  & 1.0(2) & 0.58557(1) & 0.586(24) & -0.4(4) & 0.039(5) & 0.40(10) & 0.7(5) & 0.57 \\
   & 64 & 0.423(4)  & 1.0(3) & 0.58557(2) & 0.588(35) & -0.4(4) & 0.04(2)  & 0.4(2) & 0.8(8) & 0.61 \\
   & 128& 0.42(1)   & 0.01(6)& 0.58557(3) & 0.905(617)& -0.00(4) & 0.04(5) & 0.4(6) & 1.3(2) & 0.69 \\
 4 & 32 & 0.426(3)  & 1.7(6) & 0.58580(2) & 0.521(42) & -1.1(4.3) & 0.037(3) & 0.33(10) & -0.9(1.5) & 0.64 \\
   & 64 & 0.430(2)  & 1.8(5) & 0.58583(2) & 0.513(36) & -1.1(4.8) & 0.07(4) & 0.6(2) & -3.2(3.8) & 0.62 \\
   & 128& 0.429(7)  & 0.9(1.3) & 0.58582(4) & 0.598(155) & -0.2(1.4) & 0.0(1) & 0.5(9) & 2.6(5.9) & 0.55
\end{tabular}
\end{ruledtabular}
\end{table*}

\begin{table*}[t]
\caption{Representative three-dimensional wrapping-probability fits based on the directional average $R_x$ and the full form of Eq.~\eqref{eq:supp_wrapping_fit}.  The adopted thresholds in the Letter are $\pc=0.34849(3)$, $0.35832(6)$, $0.35811(5)$, and $0.35810(7)$ for $m=1,2,3,$ and $4$, respectively.}
\label{tab:wrap_3d}
\begin{ruledtabular}
\scriptsize
\setlength{\tabcolsep}{2pt}
\renewcommand{\arraystretch}{0.92}
\begin{tabular}{lllllllllll}
 $m$ & $L_{\min}$ & $R_c$ & $a_1$ & $\pc$ & $y_p$ & $a_2$ & $b_1$ & $y_1$ & $c_1$ & $\chi^2/\mathrm{DOF}$ \\
 \colrule
 1 & 16 & 0.350(1) & 1.4(1) & 0.348488(9) & 0.855(15) & 0.3(2) & 0.2(1) & 1.2(2) & -0.8(2.2) & 1.08 \\
   & 24 & 0.350(2) & 1.2(1) & 0.34849(1)  & 0.873(20) & 0.2(2) & 0.3(4) & 1.2(6) & 3.7(6.2) & 1.08 \\
   & 32 & 0.349(5) & 1.2(3) & 0.34848(2)  & 0.873(36) & 0.2(2) & 0.1(3) & 0.9(1.0) & 1.1(3.3) & 1.27 \\
 2 & 16 & 0.368(2) & 1.3(1) & 0.35832(1)  & 0.831(19) & 0.5(3) & 0.24(8) & 1.0(2) & -0.3(1.8) & 0.65 \\
   & 24 & 0.370(2) & 1.2(1) & 0.35833(1)  & 0.838(19) & 0.5(3) & 0.6(6) & 1.4(4) & 2.3(6.2) & 0.66 \\
   & 32 & 0.36(1)  & 1.0(6) & 0.35830(4)  & 0.865(86) & 0.4(4) & 0.07(7) & 0.5(7) & 0.8(1.4) & 0.68 \\
 3 & 16 & 0.372(1) & 1.4(1) & 0.35811(1)  & 0.807(16) & -0.3(3) & 0.27(4) & 0.90(7) & -1.0(1.1) & 1.29 \\
   & 24 & 0.373(2) & 1.4(1) & 0.35812(1)  & 0.811(17) & -0.3(2) & 0.4(2) & 1.0(2) & -1.0(2.0) & 1.41 \\
   & 32 & 0.375(2) & 1.4(1) & 0.35813(2)  & 0.814(19) & -0.3(2) & 0.7(6) & 1.2(3) & -1.1(4.9) & 1.24 \\
   & 48 & 0.373(8) & 1.2(3) & 0.35812(4)  & 0.838(45) & -0.2(2) & 0.3(7) & 0.9(10) & 2.3(6.5) & 1.22 \\
 4 & 16 & 0.358(9) & 1.6(5) & 0.35802(4)  & 0.781(48) & 0.4(6) & 0.17(1) & 0.48(9) & -1.0(1.2) & 1.42 \\
   & 24 & 0.373(6) & 1.4(3) & 0.35809(4)  & 0.799(43) & 0.4(5) & 0.3(1) & 0.8(2) & -1.2(2.3) & 1.14 \\
   & 32 & 0.37(1)  & 1.1(4) & 0.35810(7)  & 0.846(66) & 0.2(4) & 0.3(3) & 0.9(5) & 1.5(3.1) & 1.32
\end{tabular}
\end{ruledtabular}
\end{table*}

\begin{figure}[t]
\centering
\includegraphics[width=0.48\textwidth]{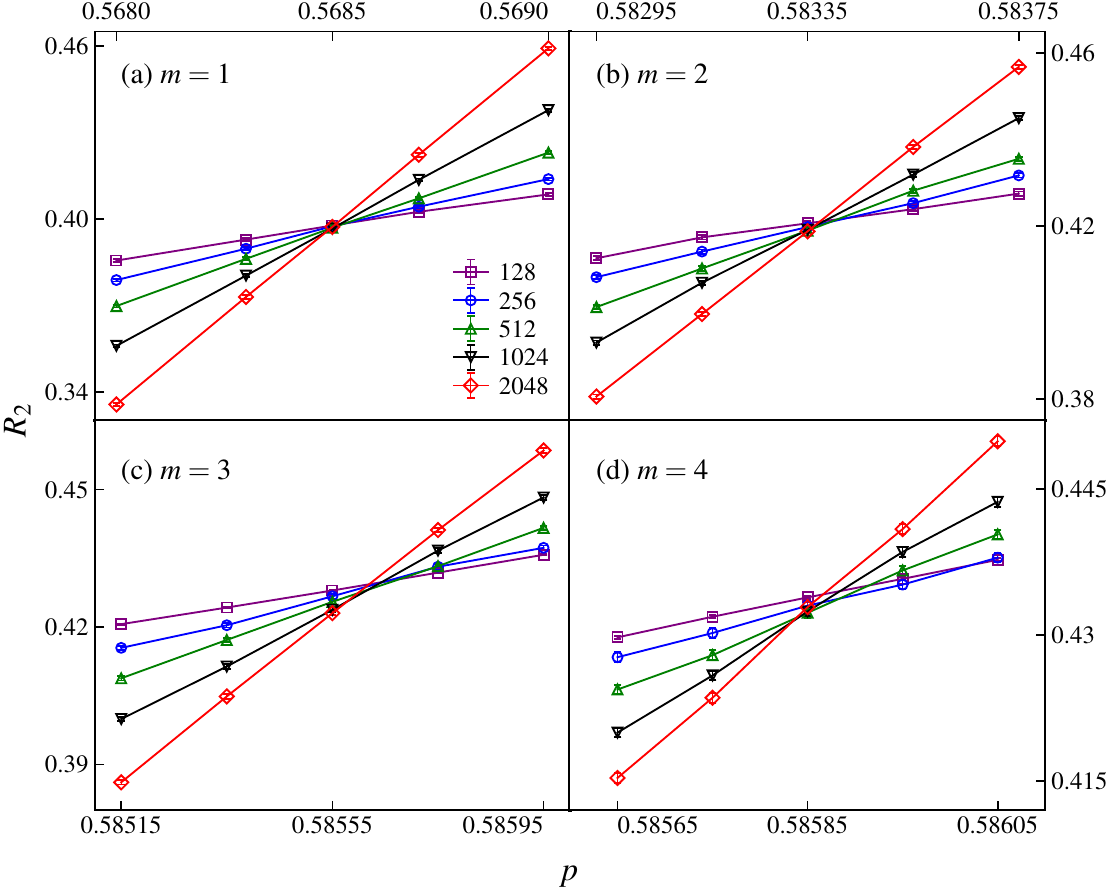}
\caption{Two-dimensional wrapping probability $R_2$ versus the final-step insertion probability for generations $m=1,\ldots,4$.  Curves for the indicated system sizes cross in a narrow interval at every generation.}
\label{fig:supp_wrapping_2d}
\end{figure}

\begin{figure}[t]
\centering
\includegraphics[width=0.48\textwidth]{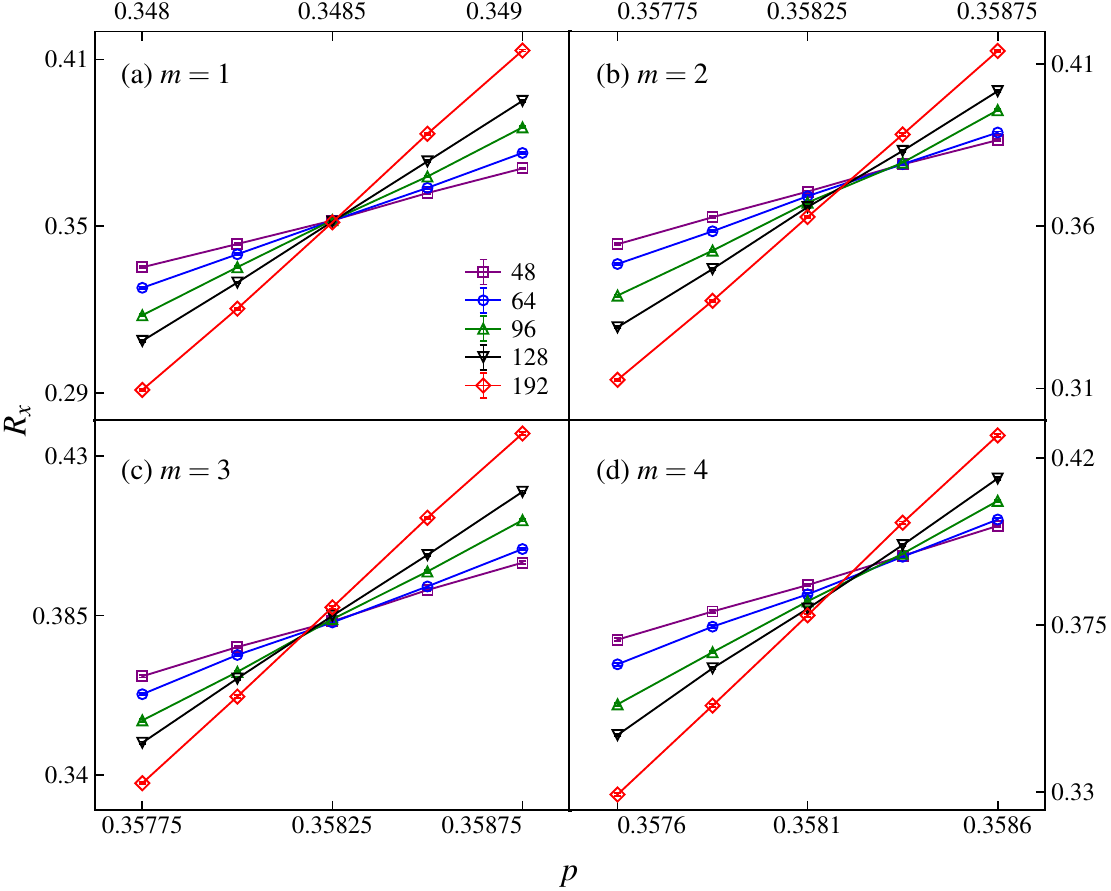}
\caption{Three-dimensional directionally averaged wrapping probability $R_x$ versus the final-step insertion probability for generations $m=1,\ldots,4$.  Curves for the indicated system sizes cross in a narrow interval at every generation.}
\label{fig:supp_wrapping_3d}
\end{figure}

Figures~\ref{fig:supp_wrapping_2d} and \ref{fig:supp_wrapping_3d} show the complete scan data underlying the two- and three-dimensional fits.  Each panel contains the raw wrapping curves for one generation and all displayed system sizes.

\section{Geometric and connectivity observables}
\label{sec:geometric_observables}

The bond-cluster observables are measured at the adopted critical points.  Let $C_1$ be the mass, namely the number of sites, of the largest bond cluster in $A_m$.  After independently coloring the bond clusters of $A_m$, we also measure the mass $S_1$ of the largest geometric spin cluster, namely a connected component of nearest-neighbor sites having the same auxiliary spin.  This observable is constructed from the colored site configuration and is distinct from the bond-cluster ensemble used to generate it.

For an observable $O$ with fractal dimension $d_O$, we use the finite-size form
\begin{equation}
 O(L)=L^{d_O}\left(a_0+a_1L^{-\omega}\right).
 \label{eq:supp_geometric_fit}
\end{equation}
In particular, $C_1\sim L^{d_{\rm FK}}$ and $S_1\sim L^{d_{\rm spin}}$.  The same form is used for the full hull length, the external-perimeter length, the number of red bonds, and the backbone mass, defining $d_H$, $d_{\rm EP}$, $d_{\rm red}$, and $d_B$, respectively.  The external perimeter is measured as the hull of the associated dense cluster, constructed by filling every nearest-neighbor bond whose endpoints lie in the same bond cluster.  We identify the backbone using the algorithm of Ref.~\cite{FangKeZhongDeng2022}.

For the effective-$\kappa$ estimates in Fig.~3(b) of the Letter, we use Theorem~1.1 of Ref.~\cite{LiuSunYuZhuang2024} for $d_{\rm spin}$ at equal coloring probability $r=1/2$.  For $d_B$, we use the backbone relation in the final equation of Ref.~\cite{NolinQianSunZhuang2024}.

Red bonds are identified without separately deleting every occupied bond.  An unwrapped depth-first-search tree assigns winding vectors to the cycles generated by non-tree edges, and their accumulation over the tree substructures identifies the bonds whose removal eliminates every nonzero winding of the cluster; thus a red bond is pivotal for the existence of any torus wrapping, rather than for wrapping in one prescribed direction.

\begin{figure}[t]
\centering
\includegraphics[width=0.48\textwidth]{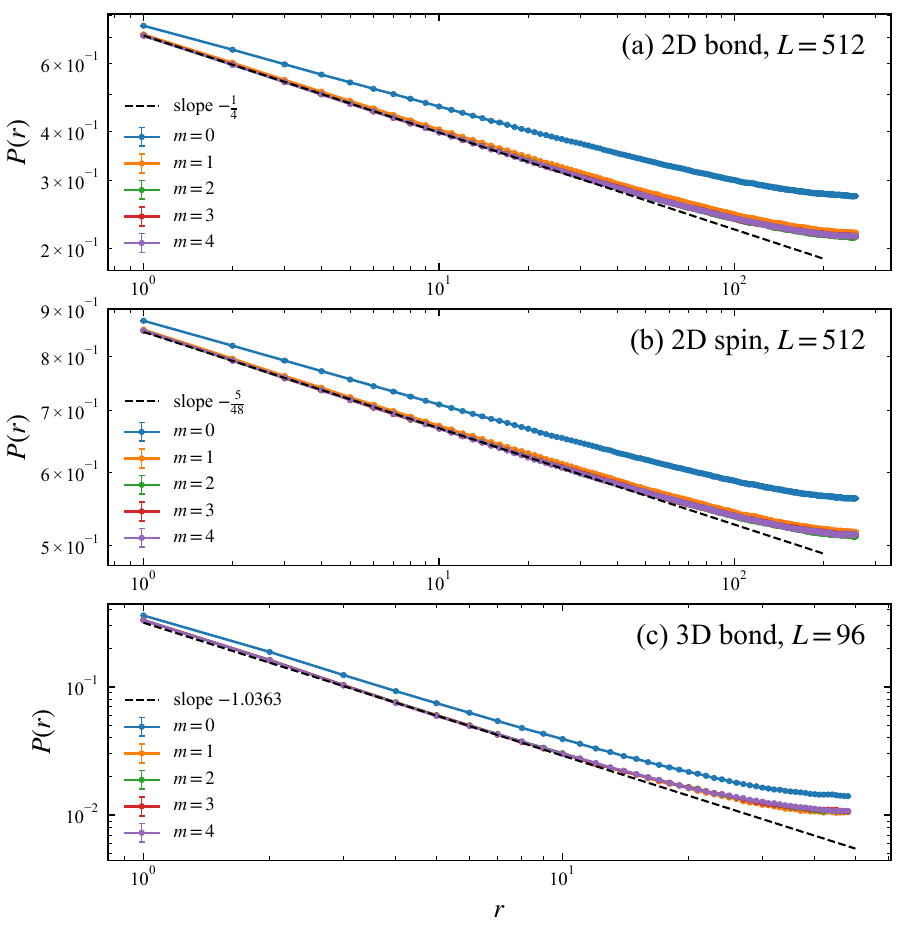}
\caption{Pair-connectivity functions for bicolored recursive percolation.  Panels (a) and (b) show the two-dimensional FK bond and geometric spin connectivities, respectively, at $L=512$, while panel (c) shows the three-dimensional FK bond connectivity at $L=96$.  The dashed lines indicate the corresponding critical FK-Ising or Ising reference power laws.}
\label{fig:supp_connectivity}
\end{figure}

To probe the long-distance connectivity structure, we directly measure the probability that two sites at a given separation belong to the same bond cluster or to the same geometric spin cluster.  Figure~\ref{fig:supp_connectivity} shows representative pair-connectivity curves.  Their power-law decay provides a direct visual comparison of the bond and spin sectors with the corresponding FK-Ising and Ising behavior.

The quoted uncertainties of scaling dimensions are statistical uncertainties from the finite-size fitting procedures only.  They do not include uncertainty propagated from the critical probabilities used to generate the preceding recursive generations.  The total uncertainties can therefore be larger than the reported fit errors.

\setcounter{section}{4}
\section{Conditional-locality analysis}
\label{sec:conditional_locality}

The colored configurations immediately before the bond-insertion step are used to test the one-site conditional spin law.  For every site $i$, define the nearest-neighbor field
\begin{equation}
 h_i=\sum_{j\in \mathrm{NN}(i)}\sigma_j,
 \label{eq:supp_nearest_field}
\end{equation}
where NN contains the four nearest neighbors in two dimensions and the six nearest neighbors in three dimensions.  The next-nearest-neighbor (NNN) field is
\begin{equation}
 d_i=\sum_{j\in \mathrm{NNN}(i)}\sigma_j,
 \label{eq:supp_nnn_field}
\end{equation}
where NNN comprises the four diagonal neighbors in two dimensions and the twelve face-diagonal neighbors in three dimensions.  Thus, the analysis retains only the nearest-neighbor and NNN fields; all remaining exterior-spin information is averaged over.

For each generation we accumulate the counts $N_+(h,d)$ and $N_-(h,d)$ of sites with $\sigma_i=+1$ and $-1$ in each $(h_i,d_i)$ sector.  The conditional log odds are estimated from pooled counts as
\begin{equation}
 \ell_m(h,d)=\log\frac{N_+(h,d)}{N_-(h,d)},
 \label{eq:supp_log_odds}
\end{equation}
The marginal log odds $\ell_m(h,L)$ are obtained by summing the counts over $d$.  The residual NNN dependence is then
\begin{equation}
 \Delta_m(h,d)=\ell_m(h,d)-\ell_m(h).
 \label{eq:supp_delta}
\end{equation}
For a nearest-neighbor Ising specification, $\ell(h,d)=2Kh$ and $\Delta(h,d)=0$.  \revision{Rather than fitting all values of the nearest-neighbor field $h$ to one finite-generation coupling, we retain the separate estimates}
\begin{equation}
 K_{\mathrm{eff},m}(h,L)=\frac{\ell_m(h,L)}{2h},
 \qquad h\ne0.
 \label{eq:supp_keff}
\end{equation}
\revision{The two-dimensional analysis uses $h=\pm2,\pm4$, and the three-dimensional analysis uses $h=\pm2,\pm4,\pm6$.  The signed values are listed separately in Tables~\ref{tab:keff-2d} and \ref{tab:keff-3d}.  For the graphical summaries only, the $+h$ and $-h$ values at fixed $|h|$ are combined with inverse-variance weights fixed from the complete sample.}

\begin{figure}[t]
\centering
\includegraphics[width=0.48\textwidth]{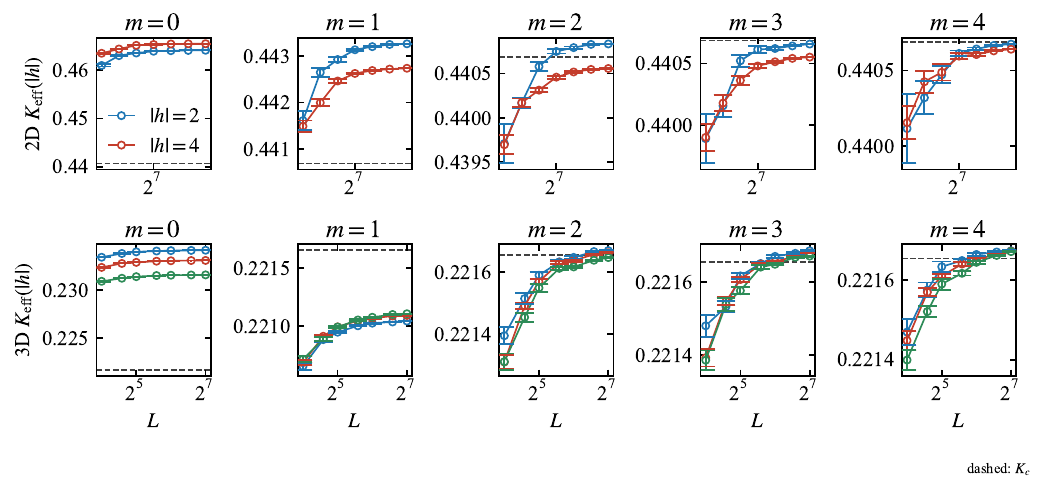}
\caption{\revision{Effective couplings $K_{\mathrm{eff},m}(|h|,L)$ evaluated separately at the available nearest-neighbor fields.  The dashed lines mark the critical Ising coupling in the corresponding dimension.  The $+h$ and $-h$ estimates are combined only for this compact visualization; their signed values are reported in Tables~\ref{tab:keff-2d} and \ref{tab:keff-3d}.  Error bars are delete-one-run jackknife errors.}}
\label{fig:supp_keff_fields}
\end{figure}

\begin{figure}[b]
\centering
\includegraphics[width=0.48\textwidth]{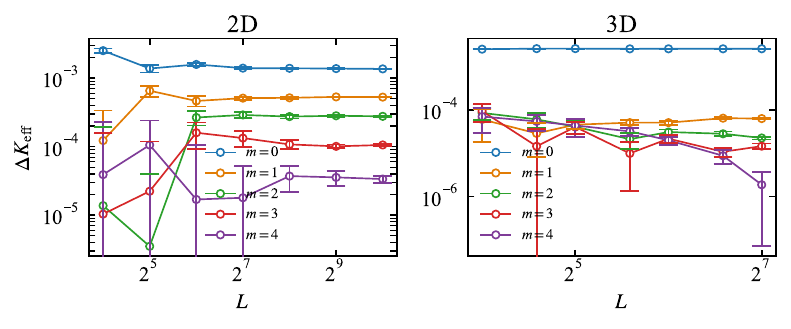}
\caption{\revision{Spread $\Delta K_{\mathrm{eff},m}(L)$ between the sign-combined estimates at different nearest-neighbor fields as a function of system size.  Each spread and its uncertainty are evaluated within the joint delete-one-run jackknife, retaining the correlations among these estimates.}}
\label{fig:supp_keff_spread}
\end{figure}

\begin{figure}[t]
\centering
\includegraphics[width=0.48\textwidth]{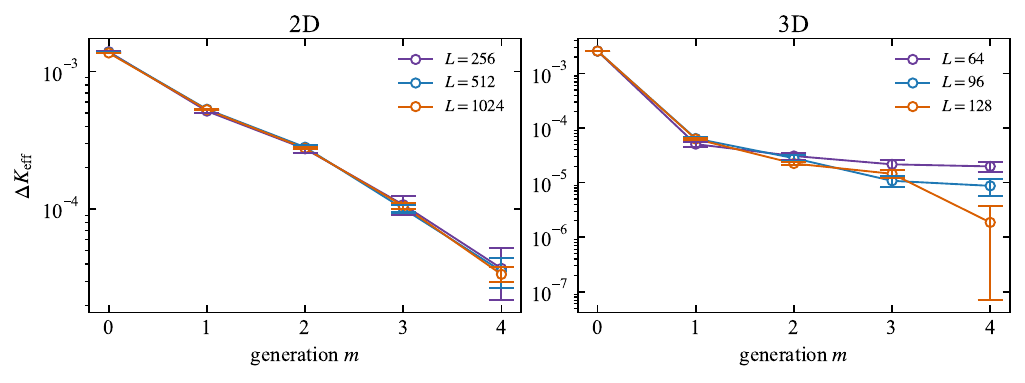}
\caption{\revision{Generation dependence of $\Delta K_{\mathrm{eff}}$ for the larger system sizes in two and three dimensions.  The pronounced suppression across successive recursion generations is visible in both cases.  Error bars are obtained by recomputing the full maximum-minus-minimum statistic in every delete-one-run sample.}}
\label{fig:supp_keff_depth}
\end{figure}

\revision{The residual field dependence is summarized by}
\begin{equation}
 \Delta K_{\mathrm{eff},m}(L)=
 \max_{|h|}K_{\mathrm{eff},m}(|h|,L)-
 \min_{|h|}K_{\mathrm{eff},m}(|h|,L).
 \label{eq:supp_keff_spread}
\end{equation}
\revision{Here $K_{\mathrm{eff},m}(|h|,L)$ denotes the sign-combined value used in the figures.  The full maximum-minus-minimum statistic is reevaluated in every delete-one-run jackknife replica, so the uncertainty of $\Delta K_{\mathrm{eff},m}(L)$ retains the correlations among the estimates at different $|h|$.  At a fixed finite generation, these estimates remain separated over the accessible sizes, whereas increasing the generation strongly suppresses their separation.}

As a complementary summary of the NNN dependence, we calculate the conditional mutual information
\begin{equation}
I_m(\sigma_i;d_i\mid h_i)=
\sum_{\sigma,h,d}P_m(\sigma,h,d)
\log\frac{P_m(\sigma,d\mid h)}{P_m(\sigma\mid h)P_m(d\mid h)}.
 \label{eq:supp_cmi}
\end{equation}
Both values of $\sigma_i$ contribute to Eq.~\eqref{eq:supp_cmi}.  The estimator is evaluated from the same pooled $(h,d,\sigma)$ counts used for Eq.~\eqref{eq:supp_log_odds}.  Finite sampling can produce a small positive estimate even when the conditional mutual information vanishes; accordingly, values close to zero are treated as a numerical resolution limit unless they exceed the resampling uncertainty.

Independent simulation runs are kept separate until the resampling stage.  The errors of $\ell_m(h,L)$, $\Delta_m(h,d)$, and $I_m(\sigma_i;d_i\mid h_i)$ are calculated with a delete-one-run jackknife.  If $\widehat O_{(a)}$ is the estimate with run $a$ omitted from $n$ independent runs, the reported standard error is
\begin{equation}
 \delta O=\left[\frac{n-1}{n}
 \sum_{a=1}^{n}\bigl(\widehat O_{(a)}-\overline O\bigr)^2\right]^{1/2},
 \qquad
 \overline O=\frac1n\sum_{a=1}^{n}\widehat O_{(a)}.
 \label{eq:supp_jackknife}
\end{equation}
The locality analysis uses generations $m=0,\ldots,4$.  In two dimensions, data are collected for $L=16,32,64,128,256,512,$ and $1024$ from twenty independent runs; in three dimensions, the corresponding sizes are $L=16,24,32,48,64,96,$ and $128$ with twenty independent runs.  For every $(L,m)$ point, the pooled counts comprise $2\times10^5L^d$ site observations.

The resulting conditional statistics reproduce the evolution shown in the Letter.  \revision{At generation zero, the log odds have a visible NNN dependence, the $K_{\mathrm{eff},m}(h,L)$ estimates obtained at different $h$ are distinct, and the conditional mutual information is appreciable.  Under the recursion, the NNN dependence and the spread in Eq.~\eqref{eq:supp_keff_spread} decrease strongly.}  At the largest sizes, the empirical conditional mutual information changes from $1.04131(44)\times10^{-4}$ at $m=0$ to $3.75(76)\times10^{-10}$ at $m=4$ in two dimensions, and from $9.256(3)\times10^{-5}$ to $1.43(24)\times10^{-10}$ in three dimensions.  The late-generation values are close to the finite-sample resolution of the aggregate-count estimator.  The delete-one-run jackknife estimates the statistical uncertainty from run to run.  Because mutual information estimated from finite counts is slightly positive even for conditionally independent variables, however, this uncertainty does not correct the small positive finite-sample offset of the estimator.

\revision{Figure~\ref{fig:supp_keff_fields} shows all available $K_{\mathrm{eff},m}(|h|,L)$ values, while Fig.~\ref{fig:supp_keff_spread} gives their size dependence.  Figure~\ref{fig:supp_keff_depth} displays the same spread across recursion generations for the larger system sizes.  All three figures and the signed tables are calculated directly from the count data retained separately for each independent run.}

\clearpage
\onecolumngrid
\section{\revision{Effective couplings at different nearest-neighbor fields}}
\label{sec:keff_tables}

\revision{The following tables give every signed $K_{\mathrm{eff},m}(h,L)$ value used in the locality analysis.  Parenthesized uncertainties refer to the final displayed digits.}

\begin{longtable}{l l l l l l}
\caption{Effective couplings at positive and negative nearest-neighbor fields in 2D. The parenthesized uncertainties are delete-one-run jackknife errors, quoted in the final displayed digits.}\label{tab:keff-2d}\\
\toprule
$m$ & $L$ & $K_{\rm eff}(h=+4)$ & $K_{\rm eff}(h=+2)$ & $K_{\rm eff}(h=-2)$ & $K_{\rm eff}(h=-4)$ \\
\midrule
\endfirsthead
\toprule
$m$ & $L$ & $K_{\rm eff}(h=+4)$ & $K_{\rm eff}(h=+2)$ & $K_{\rm eff}(h=-2)$ & $K_{\rm eff}(h=-4)$ \\
\midrule
\endhead
0 & 16 & $0.4635(2)$ & $0.4606(3)$ & $0.4612(3)$ & $0.4634(2)$ \\
0 & 32 & $0.4644(1)$ & $0.4628(2)$ & $0.4631(1)$ & $0.4643(1)$ \\
0 & 64 & $0.46512(5)$ & $0.46353(7)$ & $0.46350(9)$ & $0.46508(5)$ \\
0 & 128 & $0.46528(2)$ & $0.46387(3)$ & $0.46389(4)$ & $0.46527(2)$ \\
0 & 256 & $0.46539(1)$ & $0.46399(3)$ & $0.46400(2)$ & $0.46538(1)$ \\
0 & 512 & $0.465434(7)$ & $0.46404(2)$ & $0.46406(2)$ & $0.465423(5)$ \\
0 & 1024 & $0.465444(6)$ & $0.46407(1)$ & $0.46410(1)$ & $0.465451(5)$ \\
\midrule
1 & 16 & $0.4416(2)$ & $0.4418(3)$ & $0.4415(2)$ & $0.4414(2)$ \\
1 & 32 & $0.4421(1)$ & $0.4426(1)$ & $0.4427(1)$ & $0.44195(8)$ \\
1 & 64 & $0.44245(7)$ & $0.44292(5)$ & $0.44295(8)$ & $0.44247(5)$ \\
1 & 128 & $0.44264(2)$ & $0.44312(3)$ & $0.44319(4)$ & $0.44262(2)$ \\
1 & 256 & $0.442694(9)$ & $0.44320(1)$ & $0.44322(2)$ & $0.44269(1)$ \\
1 & 512 & $0.442724(4)$ & $0.443258(7)$ & $0.443254(7)$ & $0.442723(5)$ \\
1 & 1024 & $0.442740(3)$ & $0.443269(4)$ & $0.443270(4)$ & $0.442742(3)$ \\
\midrule
2 & 16 & $0.4396(2)$ & $0.4399(3)$ & $0.4396(3)$ & $0.4397(2)$ \\
2 & 32 & $0.44025(7)$ & $0.4403(1)$ & $0.4400(1)$ & $0.4400(1)$ \\
2 & 64 & $0.44035(5)$ & $0.44053(5)$ & $0.44065(6)$ & $0.44029(4)$ \\
2 & 128 & $0.44044(2)$ & $0.44077(2)$ & $0.44072(3)$ & $0.44048(3)$ \\
2 & 256 & $0.44051(1)$ & $0.44079(1)$ & $0.44079(2)$ & $0.44052(1)$ \\
2 & 512 & $0.440551(5)$ & $0.440840(9)$ & $0.440822(8)$ & $0.440543(6)$ \\
2 & 1024 & $0.440556(4)$ & $0.440834(3)$ & $0.440837(5)$ & $0.440560(3)$ \\
\midrule
3 & 16 & $0.4400(2)$ & $0.4399(3)$ & $0.4399(2)$ & $0.4398(1)$ \\
3 & 32 & $0.44021(9)$ & $0.4402(1)$ & $0.4401(1)$ & $0.4401(1)$ \\
3 & 64 & $0.44036(4)$ & $0.44047(7)$ & $0.44056(6)$ & $0.44037(6)$ \\
3 & 128 & $0.44049(3)$ & $0.44059(4)$ & $0.44063(3)$ & $0.44048(2)$ \\
3 & 256 & $0.44052(1)$ & $0.44061(2)$ & $0.44063(1)$ & $0.440512(9)$ \\
3 & 512 & $0.440547(9)$ & $0.440632(8)$ & $0.440657(8)$ & $0.440542(4)$ \\
3 & 1024 & $0.440553(3)$ & $0.440658(4)$ & $0.440659(5)$ & $0.440551(4)$ \\
\midrule
4 & 16 & $0.4402(1)$ & $0.4400(3)$ & $0.4402(3)$ & $0.4400(2)$ \\
4 & 32 & $0.4404(1)$ & $0.4404(1)$ & $0.4402(1)$ & $0.4404(1)$ \\
4 & 64 & $0.44054(6)$ & $0.44042(6)$ & $0.44056(8)$ & $0.44043(6)$ \\
4 & 128 & $0.44062(3)$ & $0.44058(2)$ & $0.44067(3)$ & $0.44057(2)$ \\
4 & 256 & $0.44060(1)$ & $0.44064(1)$ & $0.44064(1)$ & $0.44060(1)$ \\
4 & 512 & $0.440621(6)$ & $0.440666(7)$ & $0.440663(7)$ & $0.440635(6)$ \\
4 & 1024 & $0.440643(3)$ & $0.440671(4)$ & $0.440675(4)$ & $0.440637(3)$ \\
\bottomrule
\end{longtable}

\begin{longtable}{l l l l l l l l}
\caption{Effective couplings at positive and negative nearest-neighbor fields in 3D. The parenthesized uncertainties are delete-one-run jackknife errors, quoted in the final displayed digits.}\label{tab:keff-3d}\\
\toprule
$m$ & $L$ & $K_{\rm eff}(h=+6)$ & $K_{\rm eff}(h=+4)$ & $K_{\rm eff}(h=+2)$ & $K_{\rm eff}(h=-2)$ & $K_{\rm eff}(h=-4)$ & $K_{\rm eff}(h=-6)$ \\
\midrule
\endfirsthead
\toprule
$m$ & $L$ & $K_{\rm eff}(h=+6)$ & $K_{\rm eff}(h=+4)$ & $K_{\rm eff}(h=+2)$ & $K_{\rm eff}(h=-2)$ & $K_{\rm eff}(h=-4)$ & $K_{\rm eff}(h=-6)$ \\
\midrule
\endhead
0 & 16 & $0.23089(4)$ & $0.23240(3)$ & $0.23348(4)$ & $0.23344(5)$ & $0.23240(3)$ & $0.23098(4)$ \\
0 & 24 & $0.23125(2)$ & $0.23282(1)$ & $0.23383(3)$ & $0.23389(3)$ & $0.23280(2)$ & $0.23127(2)$ \\
0 & 32 & $0.23144(2)$ & $0.23292(1)$ & $0.23403(1)$ & $0.23403(2)$ & $0.23293(1)$ & $0.23142(2)$ \\
0 & 48 & $0.231535(9)$ & $0.233050(8)$ & $0.23411(1)$ & $0.23411(1)$ & $0.233048(6)$ & $0.231551(8)$ \\
0 & 64 & $0.231573(5)$ & $0.233097(4)$ & $0.234150(7)$ & $0.234159(8)$ & $0.233086(5)$ & $0.231584(6)$ \\
0 & 96 & $0.231601(4)$ & $0.233117(3)$ & $0.234176(7)$ & $0.234183(8)$ & $0.233123(4)$ & $0.231604(4)$ \\
0 & 128 & $0.231615(3)$ & $0.233128(3)$ & $0.234188(5)$ & $0.234202(5)$ & $0.233132(3)$ & $0.231613(3)$ \\
\midrule
1 & 16 & $0.22077(5)$ & $0.22069(3)$ & $0.22067(5)$ & $0.22064(3)$ & $0.22068(4)$ & $0.22067(4)$ \\
1 & 24 & $0.22092(3)$ & $0.22091(1)$ & $0.22088(2)$ & $0.22089(3)$ & $0.22091(2)$ & $0.22088(2)$ \\
1 & 32 & $0.22100(1)$ & $0.22100(1)$ & $0.22094(1)$ & $0.22095(1)$ & $0.22099(1)$ & $0.22099(2)$ \\
1 & 48 & $0.221046(7)$ & $0.221050(7)$ & $0.221005(6)$ & $0.220999(9)$ & $0.221047(5)$ & $0.221065(8)$ \\
1 & 64 & $0.221076(7)$ & $0.221068(2)$ & $0.221025(4)$ & $0.221028(5)$ & $0.221067(3)$ & $0.221078(5)$ \\
1 & 96 & $0.221100(3)$ & $0.221089(3)$ & $0.221037(3)$ & $0.221036(3)$ & $0.221088(2)$ & $0.221103(4)$ \\
1 & 128 & $0.221110(2)$ & $0.221091(1)$ & $0.221045(2)$ & $0.221043(3)$ & $0.221094(1)$ & $0.221106(1)$ \\
\midrule
2 & 16 & $0.22128(4)$ & $0.22135(3)$ & $0.22140(4)$ & $0.22139(3)$ & $0.22127(4)$ & $0.22133(3)$ \\
2 & 24 & $0.22146(2)$ & $0.22151(2)$ & $0.22151(2)$ & $0.22152(2)$ & $0.22148(1)$ & $0.22145(3)$ \\
2 & 32 & $0.22155(2)$ & $0.22158(1)$ & $0.22159(2)$ & $0.22159(1)$ & $0.22156(1)$ & $0.22155(1)$ \\
2 & 48 & $0.221617(8)$ & $0.221625(8)$ & $0.221629(7)$ & $0.221641(8)$ & $0.221633(6)$ & $0.221609(7)$ \\
2 & 64 & $0.221615(5)$ & $0.221641(5)$ & $0.221644(5)$ & $0.221653(5)$ & $0.221635(4)$ & $0.221620(6)$ \\
2 & 96 & $0.221641(3)$ & $0.221660(2)$ & $0.221662(3)$ & $0.221671(3)$ & $0.221656(2)$ & $0.221635(3)$ \\
2 & 128 & $0.221646(2)$ & $0.221662(1)$ & $0.221670(2)$ & $0.221670(2)$ & $0.221664(2)$ & $0.221648(2)$ \\
\midrule
3 & 16 & $0.22139(4)$ & $0.22142(3)$ & $0.22146(3)$ & $0.22152(4)$ & $0.22135(3)$ & $0.22138(4)$ \\
3 & 24 & $0.22154(2)$ & $0.22154(2)$ & $0.22153(2)$ & $0.22153(2)$ & $0.22156(2)$ & $0.22153(3)$ \\
3 & 32 & $0.22158(1)$ & $0.22161(1)$ & $0.22162(1)$ & $0.22162(1)$ & $0.22161(1)$ & $0.22158(2)$ \\
3 & 48 & $0.221640(7)$ & $0.221644(6)$ & $0.221648(7)$ & $0.221653(6)$ & $0.221651(3)$ & $0.221643(8)$ \\
3 & 64 & $0.221647(4)$ & $0.221657(3)$ & $0.221669(3)$ & $0.221675(6)$ & $0.221655(4)$ & $0.221652(7)$ \\
3 & 96 & $0.221664(3)$ & $0.221674(3)$ & $0.221679(3)$ & $0.221678(3)$ & $0.221669(2)$ & $0.221669(2)$ \\
3 & 128 & $0.221671(2)$ & $0.2216798(9)$ & $0.221686(2)$ & $0.221685(2)$ & $0.221680(1)$ & $0.221671(2)$ \\
\midrule
4 & 16 & $0.22144(5)$ & $0.22144(2)$ & $0.22147(5)$ & $0.22147(4)$ & $0.22147(4)$ & $0.22139(3)$ \\
4 & 24 & $0.22157(2)$ & $0.22158(2)$ & $0.22157(2)$ & $0.22158(2)$ & $0.22156(2)$ & $0.22145(3)$ \\
4 & 32 & $0.22157(2)$ & $0.221595(8)$ & $0.22164(2)$ & $0.22163(1)$ & $0.221629(9)$ & $0.22161(2)$ \\
4 & 48 & $0.221625(9)$ & $0.221635(6)$ & $0.221651(7)$ & $0.221647(7)$ & $0.221645(6)$ & $0.221611(8)$ \\
4 & 64 & $0.221654(5)$ & $0.221653(4)$ & $0.221663(4)$ & $0.221666(4)$ & $0.221651(4)$ & $0.221635(5)$ \\
4 & 96 & $0.221661(4)$ & $0.221665(3)$ & $0.221673(3)$ & $0.221668(3)$ & $0.221663(2)$ & $0.221663(3)$ \\
4 & 128 & $0.221672(1)$ & $0.221675(1)$ & $0.221674(2)$ & $0.221673(2)$ & $0.221670(1)$ & $0.221672(3)$ \\
\bottomrule
\end{longtable}

\twocolumngrid

\end{document}